\documentclass[prb,twocolumn,showpacs,amsmath,amssymb]{revtex4}

\usepackage{graphicx}
\usepackage{dcolumn}
\usepackage{bm}
\usepackage{epsfig}

\begin{document}

\title{Shot Noise of Spin-Decohering Transport in Spin-Orbit Coupled Nanostructures}

\author{Ralitsa L. Dragomirova}
\author{Branislav K. Nikoli\' c}
\affiliation{Department of Physics and Astronomy, University
of Delaware, Newark, DE 19716-2570, USA}

\begin{abstract}
We generalize the scattering theory of quantum shot noise to include the full spin-density matrix of electrons injected from a spin-filtering or ferromagnetic electrode into a quantum-coherent nanostructure governed by various spin-dependent  interactions. This formalism yields the spin-resolved shot noise power for different experimental measurement setups---with ferromagnetic source and ferromagnetic or normal drain electrodes---whose evaluation for the diffusive multichannel quantum wires with the Rashba (SO) spin-orbit coupling shows how spin decoherence and dephasing lead to substantial enhancement of charge current fluctuations (characterized by Fano factors $> 1/3$). However, these processes and the corresponding shot noise  increase are suppressed in narrow wires, so that  charge transport experiments measuring the Fano factor $F_{\uparrow \rightarrow \uparrow \downarrow}$ in a ferromagnet/SO-coupled-wire/paramagnet setup also quantify the degree of phase-coherence of transported spin---we predict a one-to-one correspondence between the magnitude of the spin polarization vector and $F_{\uparrow \rightarrow \uparrow \downarrow}$. 
\end{abstract}


\maketitle

\section{Introduction}

Over the past two decades, the exploration of the shot noise accompanying charge flow through mesoscopic 
conductors has become a major tool for gathering information about microscopic mechanisms of transport 
and correlations  between charges which cannot be extracted from traditional conductance measurements.~\cite{blanter2000}  Such nonequilibrium time-dependent fluctuations arise due 
to discreetness of charge, persist down to zero temperature (in contrast to thermal fluctuations which 
vanish at $T=0$), and  require stochasticity induced by either quantum-mechanical~\cite{schomerus} backscattering of electrons within a mesoscopic (i.e., smaller than the inelastic scattering length~\cite{steinbach1996}) conductor or 
by random injection processes (as in the textbook example of Schottky vacuum diode). 

The zero-frequency shot noise power $S=2FI$ of conventional unpolarized  charge current 
with average value $I$ in two-terminal non-interacting conductors reaches its maximum (the 
Poisson limit) characterized by the Fano  factor $F=1$ when transport is determined by 
uncorrelated stochastic processes.  On the other hand, the Pauli exclusion principle correlates electron motion  and suppress the shot noise  $F<1$ of non-interacting carriers, while electron-electron  interactions can also lead to super-Poissonian $F>1$ 
noise  signatures.~\cite{leo} For example, the well-known~\cite{beenakker1992}  
and experimentally confirmed~\cite{steinbach1996} $F=1/3$ universal suppression factor 
for non-interacting quasiparticles in two-terminal diffusive conductors is determined 
by the interplay of randomness in quantum-mechanical impurity scattering and the Pauli 
blocking imposed by their Fermi statistics. 

In contrast to the wealth of information acquired on the shot noise in spin-degenerate transport, 
it is only recently that the study of {\em spin-dependent shot noise} in ferromagnet-normal systems~\cite{tserkovnyak2001}  has been initiated in two-terminal~\cite{mishchenko2003,lamacraft2004,nagaev2006}  and multiterminal structures.~\cite{belzig2004,cottet2004,egues2005} In such devices ferromagnetic sources inject 
spin-polarized charge current into a paramagnetic region with interactions which affect the spin of transported electrons. For example, it has been shown that spin-flip scattering can substantially increase the shot noise above $F=1/3$ in diffusive wires attached to two ferromagnetic electrodes with antiparallel orientation of their magnetization,~\cite{mishchenko2003} as well as in the setup with the ferromagnetic source and a normal drain (collecting  both spin species) electrodes.~\cite{lamacraft2004} Thus, the enhanced shot noise power reveals additional sources of current fluctuation when spin degeneracy is lifted and particles from spin-$\uparrow$ electron subsystem are converted into spin-$\downarrow$ electrons. The non-conservation of particles in each spin subsystem as a source of additional noise is quite analogous 
to more familiar example of fluctuations of electromagnetic radiation in random optical media due to non-conservation of the number of photons.~\cite{beenakker1998} Microscopically, spin-flips are either instantaneous events generated by the collision of electrons with magnetic impurities and spin-orbit (SO) dependent scattering off static disorder,~\cite{nagaev2006} or continuous spin precession~\cite{lamacraft2004} during electron free propagation in magnetic fields imposed either externally or induced by intrinsic SO couplings~\cite{winkler_book} [whose ``internal'' magnetic field ${\bf B}_{\rm int}({\bf p})$ is momentum dependent and spin-splits the energy bands].

In particular, crucial role played by the SO interactions in all-electrical control of spin in 
semiconductor nanostructures~\cite{zutic2004} has also ignited recent studies of their signatures 
on the shot noise.~\cite{egues2005} It has been shown that the Rashba SO coupling in a 
two-dimensional electron gas (2DEG) can modulate the Fano factor of the shot noise of unpolarized 
charge current in clean beam splitter devices.~\cite{egues2005} Moreover, the Rashba SO interaction 
is solely responsible for the non-zero noise~\cite{ossipov2006} in ballistic chaotic dots by 
introducing quantum effects into the regime where electrons otherwise follow deterministic classical trajectories characterized by~\cite{schomerus} $F=0$. 

Despite these advances, key questions for the understanding of shot noise in diffusive SO-coupled nanostructures remain unanswered: {\em What is the connection between the Fano factor and the degree of quantum coherence $|{\bf P}_{\rm detect}|$ of transported spins?} {\em How does the shot noise depend on the spin polarization vector ${\bf P}_{\rm inject}$ of injected current and its direction with respect to ${\bf B}_{\rm int}({\bf p})$?} The spin polarization vector of the detected current~\cite{nikolic_purity}  ${\bf P}_{\rm detect}$ is rotated by coherent precession, as well as shrunk $0<|{\bf P}_{\rm detect}|<1$ by the D'yakonov-Perel' (DP) spin dephasing~\cite{zutic2004,wl_spin1} due to random changes in ${\bf B}_{\rm int}({\bf p})$ after electron  scatters 
off impurities or boundaries. Such different aspects of spin dynamics could leave distinctive signatures on the shot noise.~\cite{lamacraft2004}

At low temperatures, where small enough ($\lesssim 1 \ \mu$m) conductors become phase-coherent and 
Pauli blocking renders regular injection and collection of charge carriers from the bulk electrodes, 
the scattering theory of quantum transport provides~\cite{blanter2000,buttiker1992a} the celebrated 
formula for the shot noise power in terms of the transmission eigenvalues $T_n$, 
\begin{equation}\label{eq:khlus}
S = \frac{4e^3V}{h} \sum_{n=1}^M T_n(1-T_n)
\end{equation}
where $V$ is the linear response (time-independent) bias voltage.~\cite{blanter2000} Its physical interpretation is quite transparent---in the basis of eigenchannels, which diagonalize ${\bf t} {\bf t}^\dag$ with ${\bf t}$ being  the Landauer-B\" uttiker transmission matrix, a mesoscopic structure can be viewed as a parallel circuit of  $M$ (= number of transverse propagating orbital wave functions in the leads) independent one-dimensional conductors, each characterized by the transmission probability $T_n$. To get the shot noise through disordered systems, Eq.~(\ref{eq:khlus}) has to be averaged~\cite{blanter2000} over a proper distribution of $T_n$. However, this standard route {\em becomes inapplicable} for spin-polarized injection where one has to take into account the {\em spin-density matrix} of injected electrons~\cite{nikolic_purity} and therefore perform the calculations in the basis of  transverse propagating modes of the source electrode.~\cite{lamacraft2004}

Here we address questions posed above by: (i) deriving in Sec.~\ref{sec:formalism} a generalization of the scattering matrix-based formulas for the shot noise to include both the ``direction'' of injected spins and the degree of their quantum coherence, as encoded into the spin polarization vector ${\bf P}_{\rm inject}$ which specifies the spin density matrix of the current of quantum-transported electrons~\cite{nikolic_purity} $\hat{\rho}_{\rm inject}=(1+{\bf P}_{\rm inject} \cdot \hat{\bm \sigma})/2$; (ii) explicitly connecting in Secs.~\ref{sec:fano} the value of the Fano factor in the right electrode to the degree of quantum coherence of transported spin $|{\bf P}_{\rm detect}|$ extracted from recently developed~\cite{nikolic_purity} scattering approach to its spin density matrix. This formalism is applied to Rashba SO coupled quantum wires of different widths (where 
confinement affects the degree of transported spin coherence~\cite{holleitner2006a}) introduced in Sec.~\ref{sec:rashba}, and its principal results are contrasted with related spin-dependent shot noise studies in two-terminal setups in Sec.~\ref{sec:discussion}. We conclude in Sec.~\ref{sec:conclusion}.

\section{Scattering approach to spin-dependent shot noise} \label{sec:formalism}

The analysis of the spin-dependent shot noise requires to evaluate correlations between spin-resolved charge 
currents $I^\uparrow$ and $I^\downarrow$ due to the flow of spin-$\uparrow$ and spin-$\downarrow$ electrons through the terminals of a nanostructure~\cite{sauret2004}
\begin{equation}\label{eq:noise_time}
S_{\alpha \beta}^{\sigma \sigma^\prime} (t-t^\prime) = \frac{1}{2} \langle \delta \hat{I}_\alpha^\sigma(t)  
\delta \hat{I}_\beta^{\sigma^\prime}(t^\prime) +  \delta \hat{I}_\beta^{\sigma^\prime}(t^\prime)  \delta \hat{I}_\alpha^\sigma(t) \rangle.
\end{equation}
Here $\hat{I}_\alpha^\sigma(t)$ is the quantum-mechanical operator of the spin-resolved ($\sigma=\uparrow,\downarrow$) charge current in lead $\alpha$, $\delta \hat{I}_\alpha(t) = \hat{I}_\alpha(t) - \langle \hat{I}_\alpha (t) \rangle$, and $\langle \ldots \rangle$ stands for both quantum-mechanical and statistical averaging over the states in the macroscopic reservoirs to which a mesoscopic conductor is attached via semi-infinite interaction-free leads. The spin-resolved noise power between terminals $\alpha$ and $\beta$ is (conventionally defined~\cite{blanter2000} as twice) the Fourier transform of Eq.~(\ref{eq:noise_time}), $S_{\alpha\beta}^{\sigma \sigma^\prime} (\omega) =  \int d(t-t^\prime)\, e^{-i\omega(t-t^\prime)} S_{\alpha\beta}^{\sigma \sigma^\prime} (t-t^\prime)$.
The noise power of the total charge current $I_\alpha=I_\alpha^\uparrow + I_\alpha^\downarrow$ is then given by $S_{\alpha\beta}(\omega) = S_{\alpha\beta}^{\uparrow \uparrow}(\omega) + S_{\alpha\beta}^{\downarrow \downarrow}(\omega) + S_{\alpha\beta}^{\uparrow \downarrow}(\omega) + S_{\alpha\beta}^{\downarrow \uparrow}(\omega)$. 

The  scattering theory of quantum transport gives  for the operator of {\em spin-resolved} charge current of spin-$\sigma$ electrons flowing through terminal $\alpha$ 
\begin{eqnarray}\label{eq:current_operator}
\hat{I}_{\alpha}^{\sigma}(t) & = & \frac{e}{h} \sum_{n=1}^{M} \int \!\! \int dE\,dE' \, e^{i(E-E')t/\hbar} [ \hat{a}_{\alpha n}^{\sigma \dagger}(E)\hat{a}_{\alpha n}^{\sigma}(E') \nonumber \\ 
&& -\hat{b}_{\beta n}^{\sigma \dagger} (E) \hat{b}_{\beta n}^{\sigma}(E')]  
\end{eqnarray}
where the operator $\hat{a}^{\sigma \dagger}_{\alpha n}(E)$ [$\hat{a}^{\sigma}_{\alpha n}(E)$] creates [annihilates] incoming electrons in lead $\alpha$ which have energy $E$, spin-$\sigma$, and orbital part of their wave function is the transverse propagating mode $|n\rangle$. Similarly, $\hat{b}^{\sigma \dagger}_{\alpha n}$, $\hat{b}_{\alpha n}^\sigma$ denote spin-$\sigma$ electrons in the outgoing states. Using this expression in Eq.~(\ref{eq:noise_time}), and taking its Fourier transform, yields the following formula for the spin-resolved  noise power spectrum
\begin{widetext}
\begin{eqnarray}\label{eq:noise_power}
S_{\alpha\beta}^{\sigma\sigma'} (\omega) &  = & \frac{e^{2}}{h}\int dE \, \sum_{\gamma,\gamma'} \sum_{\rho,\rho'=\uparrow,\downarrow} {\rm Tr} \,  \left [{\bf A}_{\gamma\gamma'}^{\rho\rho'}(\alpha,\sigma,E,E+\hbar\omega){\bf A}_{\gamma'\gamma}^{\rho'\rho}(\beta,\sigma',E+\hbar\omega,E) \right]  \nonumber \\ 
&& \times \{f_{\gamma}^{\rho}(E)[1-f_{\gamma'}^{\rho'}(E+\hbar\omega)]+f_{\gamma'}^{\rho'}(E+\hbar\omega)[1-f_{\gamma}^{\rho}(E)]\}, 
\end{eqnarray}
where $f_\gamma^{\rho}(E)$ is the Fermi function in lead $\gamma$ kept at temperature $T_\gamma$ and spin-dependent chemical potential $\mu_\gamma^{\rho}$ of spin-$\rho$ electrons ($\rho=\uparrow,\downarrow$). The B\" uttiker's current matrix~\cite{buttiker1992a} ${\bf A}_{\beta\gamma}^{\rho\rho'}(\alpha,\sigma,E,E')$, whose elements are 
\begin{eqnarray} \label{eq:amatrix}
[{\bf A}_{\beta\gamma}^{\rho\rho'}(\alpha,\sigma,E,E')]_{mn}  =  \delta_{m n} \delta_{\beta \alpha} \delta_{\gamma \alpha} \delta^{\sigma \rho} \delta^{\sigma
  \rho'}  -\sum_{k} [{\bf s}_{\alpha \beta}^{\sigma \rho\dagger}(E)]_{mk} [{\bf s}_{\alpha
  \gamma}^{\sigma \rho'}(E')]_{kn},
\end{eqnarray}
is now generalized to include explicitly spin degrees of freedom through the spin-resolved scattering matrix connecting operators $\hat{a}^{\sigma}_{\alpha n}(E)$ and $\hat{b}^{\sigma}_{\alpha n}(E)$ via $\hat{b}_{\alpha n}^{\sigma}(E)=\sum_{\beta m} [{\bf s}_{\alpha \beta}^{\sigma \sigma'}]_{nm}(E) \hat{a}_{\beta m}^{\sigma'}(E)$. In 
the zero-temperature limit the thermal (Johnson-Nyquist) contribution to the noise vanishes and the Fermi function becomes a step function $f_\alpha^{\rho}(E)=\theta(E-\mu_\alpha^{\rho})$. Evaluation of Eq.~(\ref{eq:noise_power}) for the zero-temperature and zero-frequency limit, $S_{\alpha\beta}^{\sigma \sigma^\prime} \equiv S_{\alpha\beta}^{\sigma \sigma^\prime}(\omega=0,T=0)$, in the left lead $\alpha=2=\beta$ of a two-terminal mesoscopic device yields  
our principal result---the scattering theory formula for the shot noise arising in the course of propagation of spin-polarized current through a region with spin-dependent interactions:
\begin{subequations}\label{eq:noise_resolved}
\begin{eqnarray}
S_{22}^{\uparrow\uparrow} & = & \displaystyle\frac{2e^{2}}{h}\left[  {\rm Tr} \, \left (
    {\bf t}_{21}^{\uparrow\uparrow}{\bf t}_{21}^{\uparrow\uparrow\dagger}\right)eV
    +{\rm Tr}\left({\bf t}_{21}^{\uparrow\downarrow}{\bf t }_{21}^{\uparrow\downarrow\dagger}\right)\frac{1-|{\bf P}_{\rm inject}|}{1+|{\bf P}_{\rm inject}|}eV - {\rm Tr}\, \left 
     ({\bf t}_{21}^{\uparrow\downarrow}{\bf t}_{21}^{\uparrow\downarrow\dagger}{\bf t}_{21}^{\uparrow\downarrow}{\bf t}_{21}^{\uparrow\downarrow\dagger}\right )\frac{1-|{\bf P}_{\rm inject}|}{1+|{\bf P}_{\rm inject}|}eV
     \right. \nonumber \\
    \displaystyle  &  & -\left. {\rm Tr}\left ({\bf t}_{21}^{\uparrow\uparrow}{\bf t}_{21}^{\uparrow\uparrow\dagger}{\bf t}_{21}^{\uparrow\uparrow}{\bf t}_{21}^{\uparrow\uparrow\dagger}\right ){eV} -   
      2{\rm Tr}\left ({\bf t}_{21}^{\uparrow\downarrow}{\bf t}_{21}^{\uparrow\downarrow\dagger}{\bf t}_{21}^{\uparrow\uparrow}{\bf t}_{21}^{\uparrow\uparrow\dagger}\right )\frac{1-|{\bf P}_{\rm inject}|}{1+|{\bf P}_{\rm inject}|}eV\right], \\
     S_{22}^{\downarrow\downarrow} & = & \displaystyle\frac{2e^{2}}{h}\left[  {\rm Tr}\left (
    {\bf t}_{21}^{\downarrow\downarrow}{\bf t}_{21}^{\downarrow\downarrow\dagger}\right)\frac{1-|{\bf P}_{\rm inject}|}{1+|{\bf P}_{\rm inject}|}eV
    +{\rm Tr}\left({\bf t}_{21}^{\downarrow\uparrow}{\bf t }_{21}^{\downarrow\uparrow\dagger}\right)eV - {\rm Tr}\left 
     ({\bf t}_{21}^{\downarrow\downarrow}{\bf t}_{21}^{\downarrow\downarrow\dagger}{\bf t}_{21}^{\downarrow\downarrow}{\bf t}_{21}^{\downarrow\downarrow\dagger}\right )\frac{1-|{\bf P}_{\rm inject}|}{1+|{\bf P}_{\rm inject}|}eV
     \right. \nonumber \\
\displaystyle  &  & -\left. {\rm Tr}\left ({\bf t}_{21}^{\downarrow\uparrow}{\bf t}_{21}^{\downarrow\uparrow\dagger}{\bf t}_{21}^{\downarrow\uparrow}{\bf t}_{21}^{\downarrow\uparrow\dagger}\right ){eV} -      
           2{\rm Tr}\left ({\bf t}_{21}^{\downarrow\uparrow}{\bf t}_{21}^{\downarrow\uparrow\dagger}{\bf t}_{21}^{\downarrow\downarrow}{\bf t}_{21}^{\downarrow\downarrow\dagger}\right )\frac{1-|{\bf P}_{\rm inject}|}{1+|{\bf P}_{\rm inject}|}eV\right], \\
S_{22}^{\uparrow\downarrow} & = & -\displaystyle\frac{2e^{2}}{h}\left[{\rm Tr}\left (
    {\bf t}_{21}^{\downarrow\uparrow}{\bf t}_{21}^{\uparrow\uparrow\dagger}{\bf t}_{21}^{\uparrow\downarrow}{\bf t}_{21}^{\downarrow\downarrow\dagger}\right)\frac{1-|{\bf P}_{\rm inject}|}{1+|{\bf P}_{\rm inject}|}eV
    +{\rm Tr}\left({\bf t}_{21}^{\downarrow\uparrow}{\bf t }_{21}^{\uparrow\uparrow\dagger}{\bf t}_{21}^{\uparrow\uparrow}{\bf t}_{21}^{\downarrow\uparrow\dagger}\right)eV 
     \right. \nonumber \\ 
     \displaystyle  &  & +\left.
    {\rm Tr} \left 
     ({\bf t}_{21}^{\downarrow\downarrow}{\bf t}_{21}^{\uparrow\downarrow\dagger}{\bf t}_{21}^{\uparrow\uparrow}{\bf t}_{21}^{\downarrow\uparrow\dagger}\right) \frac{1-|{\bf P}_{\rm inject}|}{1+|{\bf P}_{\rm inject}|}eV   
    + {\rm Tr}\left({\bf t}_{21}^{\downarrow\downarrow}{\bf t}_{21}^{\uparrow\downarrow\dagger}{\bf t}_{21}^{\uparrow\downarrow}{\bf t}_{21}^{\downarrow\downarrow\dagger}\right )\frac{1-|{\bf P}_{\rm inject}|}{1+|{\bf P}_{\rm inject}|}eV
     \right],\\
S_{22}^{\downarrow\uparrow} & = & -\displaystyle\frac{2e^{2}}{h}\left[{\rm Tr}\left (
    {\bf t}_{21}^{\uparrow\uparrow}{\bf t}_{21}^{\downarrow\uparrow\dagger}{\bf t}_{21}^{\downarrow\downarrow}{\bf t}_{21}^{\uparrow\downarrow\dagger}\right)\frac{1-|{\bf P}_{\rm inject}|}{1+|{\bf P}_{\rm inject}|}eV
    +{\rm Tr}\left({\bf t}_{21}^{\uparrow\uparrow}{\bf t }_{21}^{\downarrow\uparrow\dagger}{\bf t}_{21}^{\downarrow\uparrow}{\bf t}_{21}^{\uparrow\uparrow\dagger}\right)eV  
    \right. \nonumber \\ 
     \displaystyle  &  & +\left. {\rm Tr}\left 
     ({\bf t}_{21}^{\uparrow\downarrow}{\bf t}_{21}^{\downarrow\downarrow\dagger}{\bf t}_{21}^{\downarrow\downarrow}{\bf t}_{21}^{\uparrow\downarrow\dagger}\right )\frac{1-|{\bf P}_{\rm inject}|}{1+|{\bf P}_{\rm inject}|}eV 
     + {\rm Tr}\left ({\bf t}_{21}^{\uparrow\downarrow}{\bf t}_{21}^{\downarrow\downarrow\dagger}{\bf t}_{21}^{\downarrow\uparrow}{\bf t}_{21}^{\uparrow\uparrow\dagger}\right )\frac{1-|{\bf P}_{\rm inject}|}{1+|{\bf P}_{\rm inject}|}{eV}
     \right]. 
\end{eqnarray}
\end{subequations}
\end{widetext}
Here the elements of the transmission matrix ${\bf t}_{21}^{\sigma \sigma'}$, which is a block of the full scattering matrix, determine the probability $|[{\bf t}_{21}^{\sigma \sigma^\prime}]_{nm}|^2$ for spin-$\sigma^\prime$  electron incident in lead $1$ in the orbital conducting channel $|m\rangle$ to be transmitted to lead $2$ as spin-$\sigma$ electron in channel $|n\rangle$. The direction of the spin-polarization vector of injected electrons selects the spin-quantization axis for $\uparrow$, $\downarrow$, while its magnitude quantifies the degree of spin polarization which was introduced into Eq.~(\ref{eq:noise_power}) via the spin-dependent electrochemical potentials in the injecting (left) lead,~\cite{egues2005}  $\mu_1^\uparrow=E_F+eV$ and $\mu_1^\downarrow=E_F+eV(1-|{\bf P}_{\rm inject}|)/(1+|{\bf P}_{\rm detect}|)$. In the collecting (right) lead the chemical potentials for both spin-species are the same $\mu_2^\uparrow = \mu_2^\downarrow=E_F$, where $E_F$ is the Fermi energy. For instance, injection of fully spin-$\uparrow$ polarized current $|{\bf P}_{\rm inject}|=1$ from the left lead (e.g., made of half-metallic ferromagnet) means that there is no voltage drop for spin-$\downarrow$ electrons $\mu_1^\downarrow=\mu_2^\downarrow=E_F$, so that they do not contribute to transport. 

\begin{figure*}
\centerline{\psfig{file=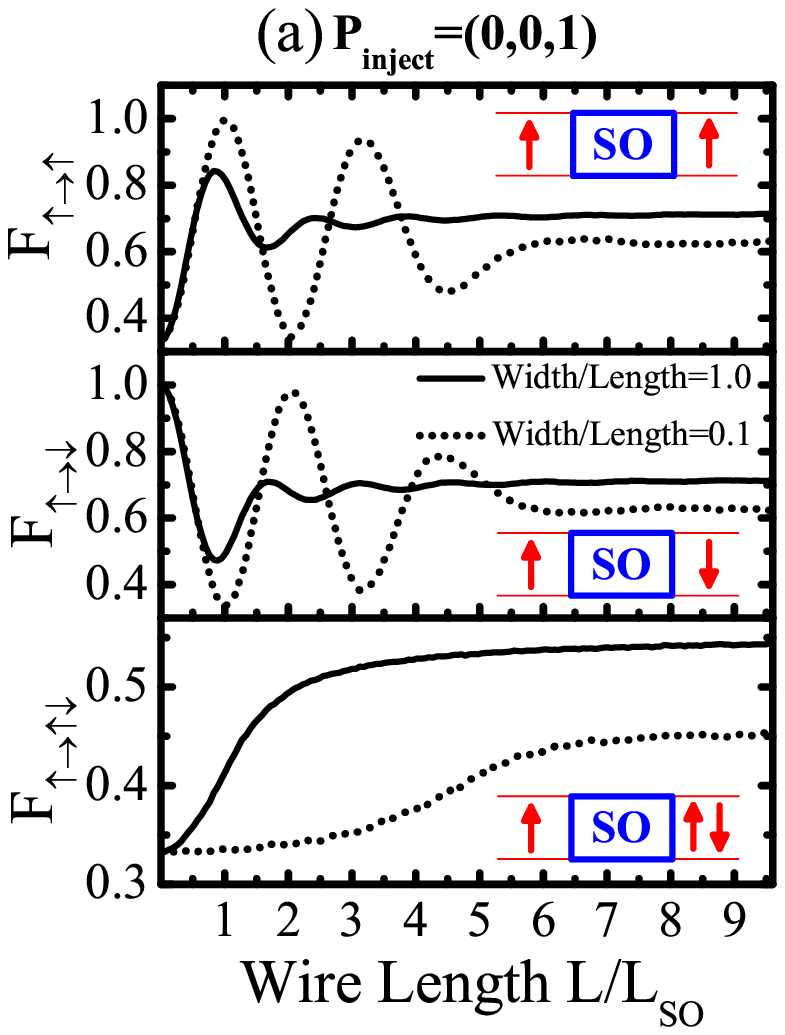,scale=0.65,angle=0} \psfig{file=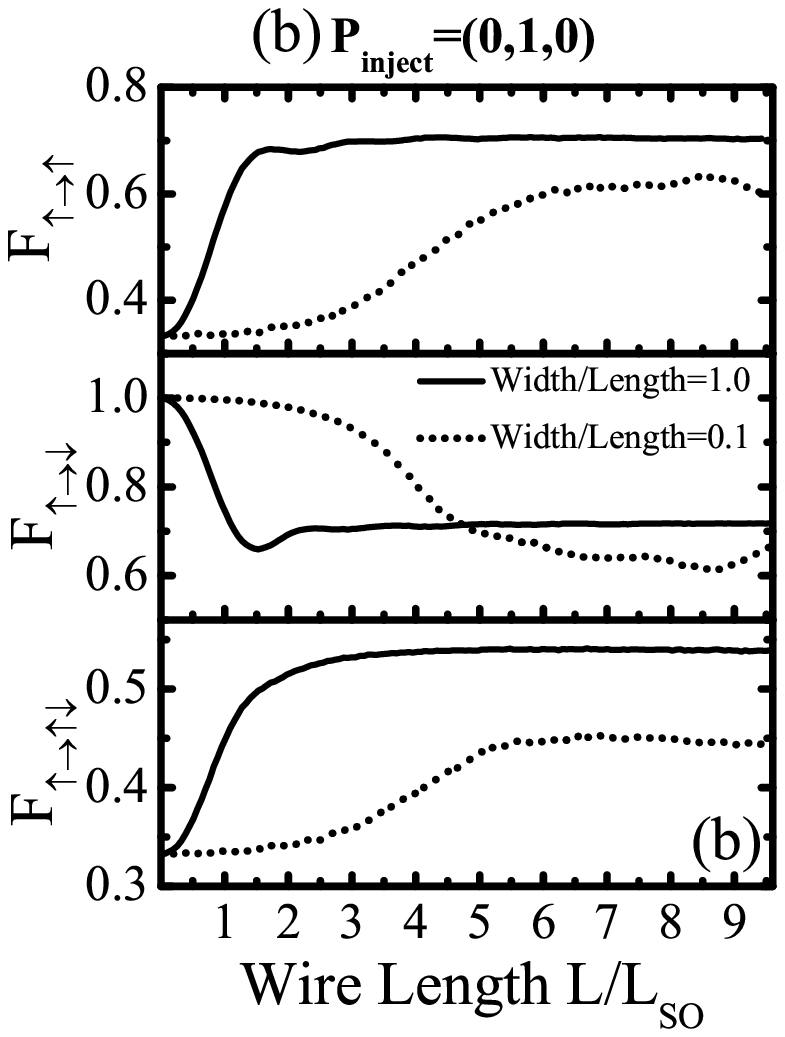,scale=0.65,angle=0} \psfig{file=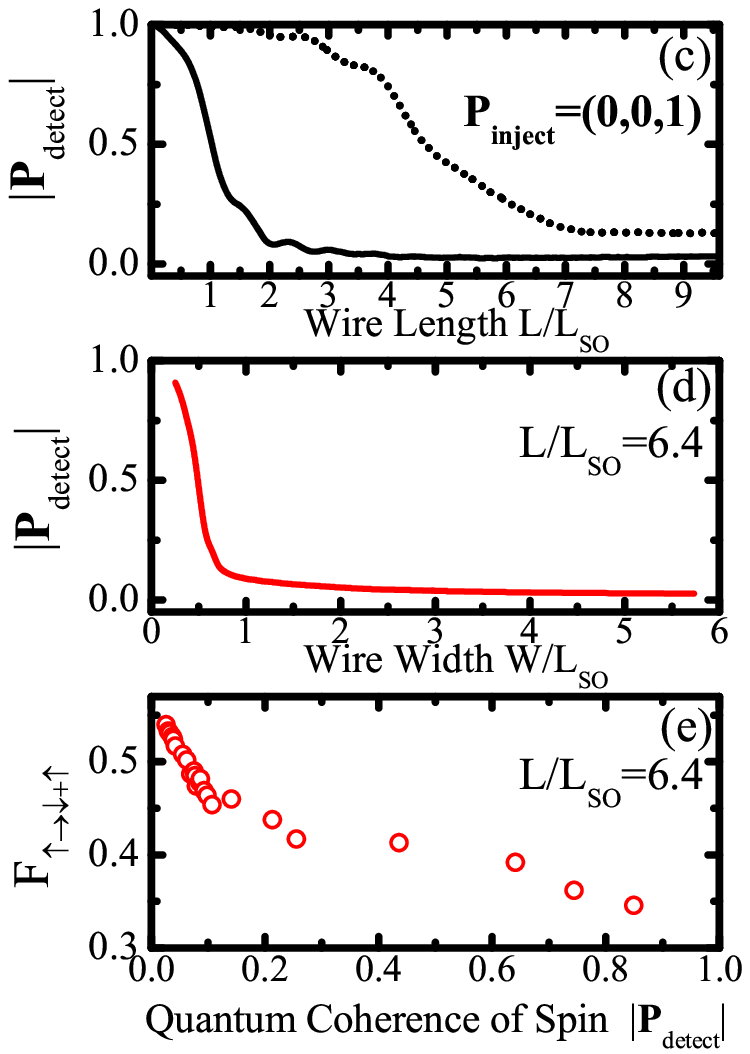,scale=0.65,angle=0}}
\caption{(Color online) Panels (a) and (b) show the Fano factor vs. the spin precession length $L_{\rm SO}$ for different two-terminal setups where 100\% spin-$\uparrow$ polarized charge current is injected from the  source electrode (e.g., half-metallic ferromagnet) into a diffusive Rashba SO-coupled wire and spin-resolved charge currents $I^\uparrow$ (top), $I^\downarrow$ (middle), or both $I^\uparrow + I^\downarrow$ (bottom), 
are collected in the drain electrode. Panel (c) shows the corresponding decay of the degree of phase-coherence of transported spin, as quantified by the magnitude of the Bloch vector of charge current, which is $|{\bf P}_{\rm inject}|=1$ (signifying fully coherent pure spin state) in the left lead and 
pointing along the $z$-axis in (a) and the $y$-axis in (b). For fixed $L$ and $L_{\rm SO}$, the decay of 
$|{\bf P}_{\rm detect}|$ is suppressed in narrow wires [panel (d)], which establishes a one-to-one 
correspondence between the Fano factor $F_{\uparrow \rightarrow \uparrow\downarrow}$ and $|{\bf P}_{\rm detect}|$ [panel (e)]. Note that the Fano factors  attaining universal value~\cite{beenakker1992,steinbach1996} $F_{\uparrow \rightarrow \uparrow} = F_{\uparrow \rightarrow \uparrow \downarrow}=1/3$ in the limit of zero SO coupling $L/L_{\rm SO} \rightarrow 0$ demonstrate that our wires are in the diffusive transport regime for selected disorder strengths.
}\label{fig:noise} 
\end{figure*}

Equations~(\ref{eq:noise_resolved}), together with the expressions for average spin-resolved currents 
collected in the right lead,
\begin{subequations}\label{eq:idetect}
\begin{eqnarray}
I_{2}^{\uparrow} \equiv \langle \hat{I}_2^\uparrow(t) \rangle & = &  \left(G_{21}^{\uparrow\uparrow}+G_{21}^{\uparrow\downarrow}\frac{1-|{\bf P}_{\rm inject}|}{1+|{\bf P}_{\rm inject}|}\right)V \\
I_{2}^{\downarrow} \equiv \langle \hat{I}_2^\downarrow(t) \rangle & = &  \left(G_{21}^{\downarrow\uparrow}+G_{21}^{\downarrow\downarrow}\frac{1-|{\bf P}_{\rm inject}|}{1+|{\bf P}_{\rm inject}|}\right)V 
\end{eqnarray}
\end{subequations}
define the Fano factors for parallel and antiparallel spin valve setups,
\begin{eqnarray}\label{eq:fano_spin_valve}
F_{\uparrow \rightarrow \uparrow} & = & \frac{S_{22}^{\uparrow \uparrow}(|{\bf P}_{\rm inject}|=1)}{2 e I_2^\uparrow(|{\bf P}_{\rm inject}|=1)}, \\ 
F_{\uparrow \rightarrow \downarrow} & = & \frac{S_{22}^{\downarrow \downarrow}(|{\bf P}_{\rm inject}|=1)}{2 e I_2^\downarrow(|{\bf P}_{\rm inject}|=1)}.
\end{eqnarray}
These equations straightforwardly also yield the Fano factor for a ferromagnet/Rashba-wire/paramagnet configuration
\begin{equation}\label{eq:fano_para}
F_{\uparrow \rightarrow \uparrow\downarrow} = \frac{S_{22}(|{\bf P}_{\rm inject}|=1)}{2 e I_2(|{\bf P}_{\rm inject}|=1)}, 
\end{equation}
where $I_2 =  I_{2}^{\uparrow} + I_{2}^{\downarrow}$ is the sum of both spin-resolved currents collected in the right paramagnetic lead. Here the spin-resolved two-terminal conductances are given by the Landauer-type formula 
\begin{equation}\label{eq:landauer}
G_{2 1}^{\sigma \sigma^\prime}=\frac{e^2}{h} \sum_{n,m=1}^{M} |[{\bf t}_{2 1}^{\sigma \sigma^\prime}]_{nm}|^2.
\end{equation}
The consequences of the scattering theory expressions, Eq.~(\ref{eq:noise_resolved}) and Eq.~(\ref{eq:idetect}), can be worked out either by analytical means (such as the random matrix theory~\cite{lamacraft2004} applicable for~\cite{aleiner2001} $L \ll L_{\rm SO}$, 
or by matching the wave functions across single- or at most two-channel structures~\cite{egues2005}) or by numerically exact real$\otimes$spin space Green functions~\cite{nikolic_purity} employed here to take as an input the microscopic Hamiltonian Eq.~(\ref{eq:rashba}) of both weekly ($L \ll L_{\rm SO}$) and strongly 
($L \ge L_{\rm SO}$) SO-coupled {\em multichannel} nanostructure. The central quantity of this formalism is 
the retarded Green function of the scattering region $\hat{G}^r=[E-\hat{H}_{\rm open}]^{-1}$ associated with the Hamiltonian $\hat{H}_{\rm open} = \hat{H} + \hat{\Sigma}_1^{r,\uparrow} + \hat{\Sigma}_1^{r,\downarrow} 
+  \hat{\Sigma}_2^{r,\uparrow} + \hat{\Sigma}_2^{r,\downarrow}$ of the open system where (non-Hermitian) retarded self-energies $\hat{\Sigma}_\alpha^{r,\sigma}$ introduced by the interaction with the leads determine escape rates of spin-$\sigma$ electrons into the electrodes. The retarded Green functions yields the spin-resolved transmission matrix through 
\begin{equation}\label{eq:transmission}
  {\bf t}_{2 1}^{\sigma \sigma^\prime}  =   2 \sqrt{-\text{Im} \, \hat{\Sigma}_2^{r,\sigma}} \cdot \hat{G}_{2 1}^{r,\sigma \sigma^\prime} \cdot \sqrt{-\text{Im}\, \hat{\Sigma}_1^{r,\sigma'}}.
\end{equation}
For simplicity, we assume that $\hat{\Sigma}^{r,\uparrow}=\hat{\Sigma}^{r,\downarrow}$, which 
experimentally corresponds to identical conditions for the injection of both spin species.

\section{Shot noise in diffusive Rashba SO-coupled wires} \label{sec:rashba}

We focus on quantum wires realized using 2DEG with the Rashba SO coupling~\cite{winkler_book} generated by structural inversion asymmetry of the semiconductor heterostructure 
hosting the 2DEG in the $xy$-plane. They are described by the effective mass Hamiltonian 
\begin{eqnarray}\label{eq:rashba}
\hat{H} & = & \frac{\hat{p}_x^2+\hat{p}_y^2}{2m^*}  + \frac{\alpha}{\hbar} \left( \hat{p}_y \otimes \hat{\sigma}_x  - \hat{p}_x \otimes \hat{\sigma}_y  \right) \nonumber \\ 
&& + V_{\rm confinement}(y) + V_{\rm disorder}(x,y).
\end{eqnarray}
Its ``internal'' magnetic field ${\bf B}_{\rm int}({\bf p})=-(2 \alpha/ g\mu_B)(\hat{\bf p} \times \hat{z})$ ($\hat{z}$ is the unit vector orthogonal to 2DEG) is nearly parallel to the transverse $y$-axis. Therefore, the injected $z$-axis polarized spins are precessing within the wires, while the $y$-axis polarized spins are in the eigenstates of the corresponding Zeeman term and do not precess.~\cite{nikolic_purity} This leads to a difference in the shot noise when changing the spin-polarization vector of the injected current in the ``polarizer-analyzer'' scheme in the top and middle panels of Figs.~\ref{fig:noise}(a) and \ref{fig:noise}(b).

Moreover, in both cases and within the asymptotic limit $L \gg L_{\rm SO}$, where $L$ is the wire length and $L_{\rm SO}$ is the spin precession length, we find the shot noise increase above the universal Fano  factor $F=1/3$ for all three measurement geometries: 

\begin{itemize}

\item spin valves with parallel magnetization of the electrodes where $\uparrow$-electrons injected from the 
left lead and $\uparrow$-electrons collected in the right lead---a situation described by the Fano factor $F_{\uparrow \rightarrow \uparrow}$, 

\item spin valves with antiparallel magnetization of the electrodes where $\uparrow$-electrons injected through a perfect Ohmic contact and  $\downarrow$-electrons collected, as described by the Fano factor $F_{\uparrow \rightarrow  \downarrow}$,

\item a setup with only one spin-selective electrode where $\uparrow$-electrons are injected and both $\uparrow$- and $\downarrow$-electrons are collected in the normal drain electrode, as described the 
Fano factor  $F_{\uparrow \rightarrow \uparrow \downarrow}$.

\end{itemize}

Note that on the $L_{\rm SO}= \pi \hbar^2/(2m^*\alpha)$ length scale spin precesses by an angle $\pi$ (i.e., the state  $|\!\! \uparrow \rangle$ evolves into $|\!\! \downarrow \rangle$), which in weakly disordered {\em bulk} systems also plays the role of characteristic length scale for the exponentially decaying spin-polarization in the DP spin dephasing.~\cite{zutic2004,wl_spin1,pareek2002} However, the asymptotic values~\cite{mishchenko2003,lamacraft2004}  of the corresponding Fano factors $F_{\sigma \rightarrow \sigma'} (L \gg L_{\rm SO},W) > F_{\uparrow \rightarrow \uparrow \downarrow} (L \gg L_{\rm SO},W) > 1/3$ are decreasing in narrow wires~\cite{holleitner2006a} of the width $W \ll L_{\rm SO}$ because DP spin dephasing can be suppressed by transverse confinement.~\cite{wl_spin1,nikolic_purity,holleitner2006a,pareek2002} Thus, Fig.~\ref{fig:noise}(e) demonstrates an exciting possibility for a novel experimental tool to quantify phase coherence of transported spin via purely electrical 
means where measurement of the Fano factor $F_{\uparrow \rightarrow \uparrow \downarrow}$ does not require 
any demanding spin selective detection in the right lead.

For very small SO coupling and, therefore, large $L_{\rm SO} \rightarrow \infty$, the Fano factors $F_{\uparrow \rightarrow \uparrow}$ and $F_{\uparrow \rightarrow \uparrow\downarrow}$ start from the universal value $F=1/3$ characterizing the diffusive unpolarized transport, and then increase toward their  asymptotic values, $F_{\uparrow \rightarrow \uparrow}(L \gg L_{\rm SO}) \approx F_{\uparrow \rightarrow \downarrow}(L \gg L_{\rm SO}) \simeq 0.7$ and $F_{\uparrow \rightarrow \uparrow\downarrow}(L \gg L_{\rm SO}) \simeq 0.55$. Such enhancement of the spin-dependent shot noise is due to spin {\em decoherence} and {\em dephasing} processes in SO-coupled structures that are reducing the off-diagonal elements of the  current spin-density matrix~\cite{nikolic_purity}  $\hat{\rho}_c$. Note that in these setups, the initial  $\hat{\rho}_c^2=\hat{\rho}_c$ describes pure injected spin states from the left lead. However, these asymptotic Fano factor values are lowered in narrow wires where transverse confinement slows down the DP spin relaxation in the picture of semiclassical spin diffusion,~\cite{wl_spin1} or reduces the size of the ``environment'' of orbital conducting channels (i.e., their number) to which the spin can entangle in fully quantum transport picture~\cite{nikolic_purity} employed to obtain $|{\bf P}_{\rm detect}|$ vs. wire width $W$ (at constant length $L$ and the Rashba SO coupling strength) in Fig.~\ref{fig:noise}(d). The geometrical confinement effects on spin coherence, which might be essential for the realization of {\em all-electrical} semiconductor spintronic devices~\cite{zutic2004} where spin is envisaged to be manipulated via SO couplings while avoiding their detrimental dephasing effects,~\cite{nikolic_purity} have been confirmed in recent optical spin detection experiments.~\cite{holleitner2006a}

The shot noise in the antiparallel configuration reaches the full Poissonian value $F_{\uparrow \rightarrow \downarrow}(L \ll L_{\rm SO}) \simeq 1$ in the limit of small SO coupling since the probability that the spin state which has huge overlap with $|\!\! \uparrow \rangle$ can enter into the right electrode with empty spin-$\uparrow$ states is minuscule. This leads to the tunneling-type~\cite{blanter2000,mishchenko2003} of shot noise where electrons propagate independently and without being correlated by the Fermi statistics. In the asymptotic limit $L \gg L_{\rm SO}$, injected spins loose their memory on a very short length scale so that $F_{\uparrow \rightarrow \downarrow}(L \gg L_{\rm SO})$ acquires the same asymptotic value as $F_{\uparrow \rightarrow \uparrow}(L \gg L_{\rm SO})$.

\begin{figure}
\centerline{\psfig{file=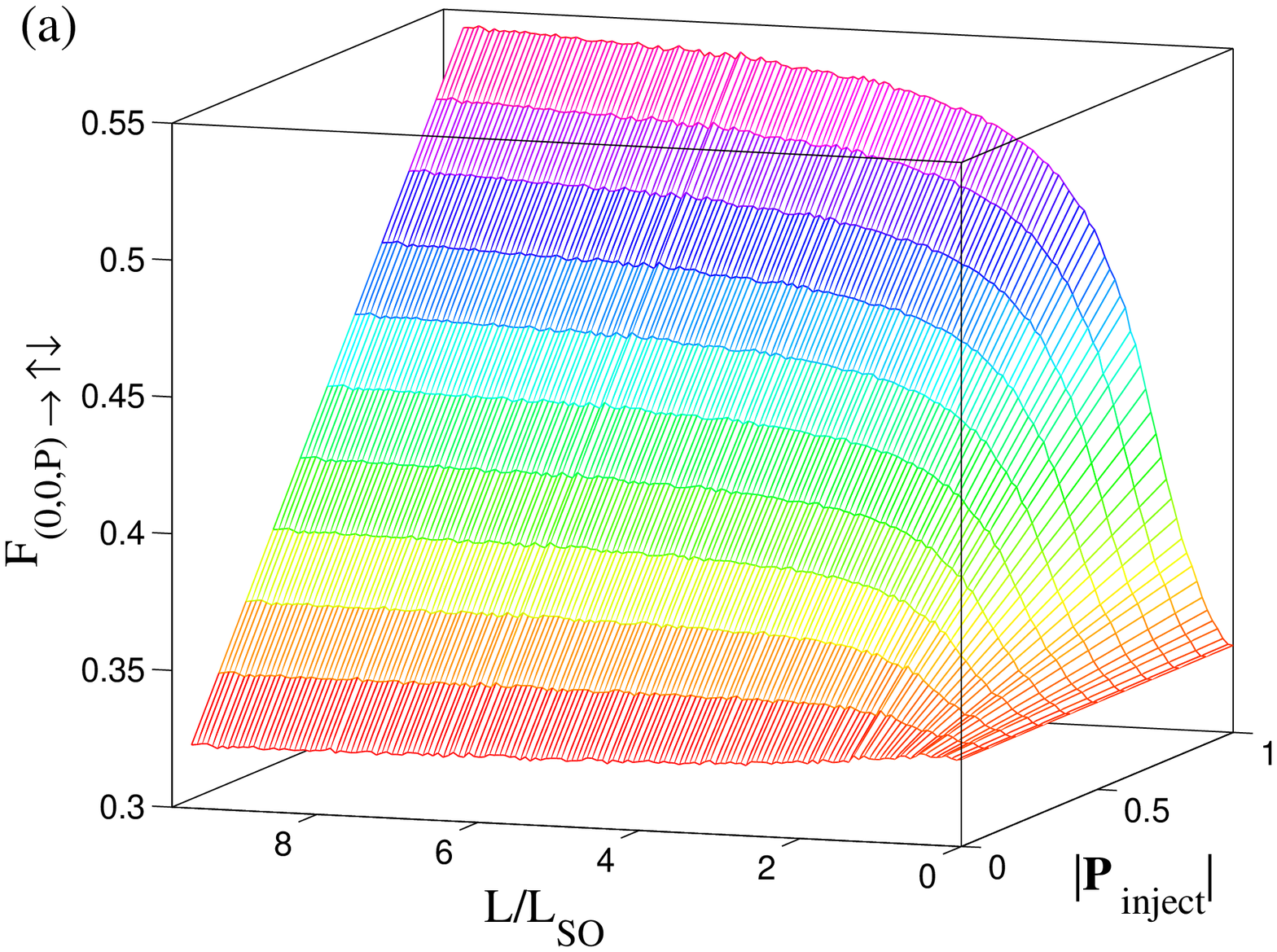,scale=0.38,angle=0}}
\centerline{\psfig{file=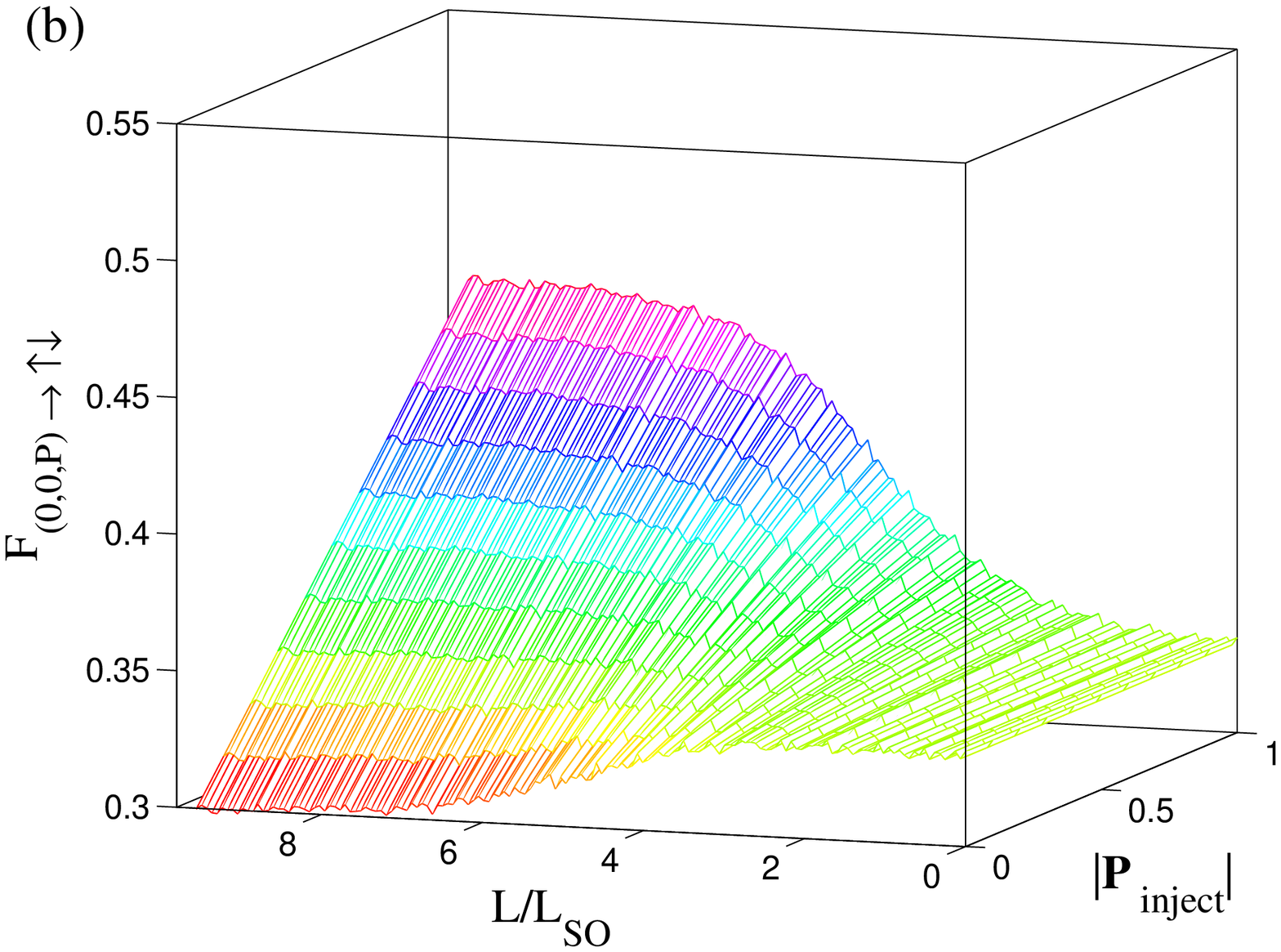,scale=0.38,angle=0}}
\caption{(Color online) Fano factor as the function of SO coupling strength $L/L_{\rm SO}$ and 
$|{\bf P}_{\rm inject}|$ for a two-terminal measuring setup where partially spin-polarized current 
[comprised of electrons with the Bloch vector ${\bf P}_{\rm inject}=(0,0,P)$] is injected from 
the source ideal lead into a diffusive Rashba SO-coupled wire and charge current of both spins 
$I^\uparrow + I^\downarrow$ is collected by the spin-nonselective drain lead. In panel (a) the 
wire length $L$ and width $W$ are the same $W/L=1$, while panel (b) plots $F_{(0,0,P) \rightarrow \uparrow\downarrow}$ in narrow wires $W/L=0.1$. Note that limiting curves extracted for $|{\bf P}_{\rm inject}|=1$ correspond to the bottom panel in Fig.~\ref{fig:noise}(a).}\label{fig:3d}
\end{figure}

Since present spintronic experiments are usually conducted by injecting partially spin-polarized charge currents $|P_{\rm inject}|<1$, we employ our general formulas Eq.~(\ref{eq:noise_resolved}) to 
obtain the Fano factor  
\begin{equation}\label{eq:fano_p}
F_{(0,0,P) \rightarrow \uparrow\downarrow} = \frac{S_{22}[{\bf P}_{\rm inject}=(0,0,P)]}{2 e I_2[{\bf P}_{\rm inject}=(0,0,P)]}, 
\end{equation}
which represents a generalization of $F_{\uparrow \rightarrow \uparrow \downarrow}$ to characterize the 
shot noise in an experimental setup where partially polarized (along the $z$-axis) electrons are injected from the left lead while both spin species are collected in the right lead. Figure 3 suggests that 
additional shot noise $F_{(0,0,P) \rightarrow \uparrow\downarrow} > 1/3$ should be observable even 
for small polarization of injected current $|{\bf P}_{\rm inject}| \equiv P \gtrsim 20\%$, in both wide and narrow SO coupled wires.

\begin{figure*}
\centerline{\psfig{file=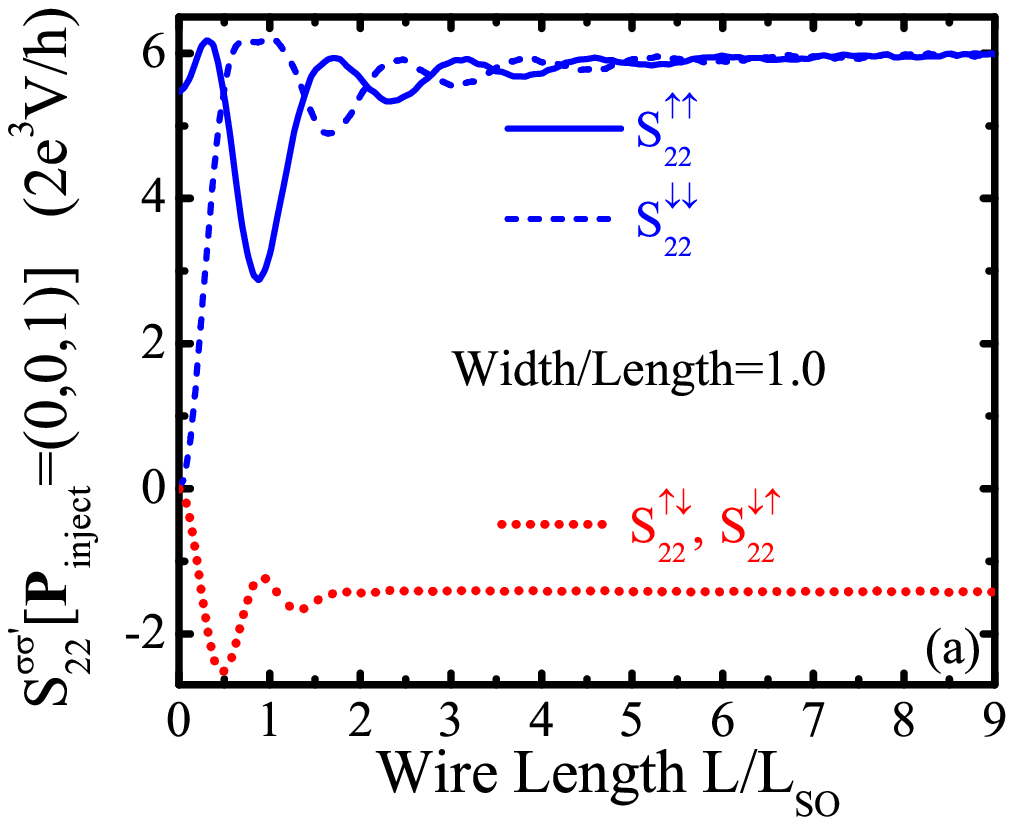,scale=0.6,angle=0} \hspace{0.5in} \psfig{file=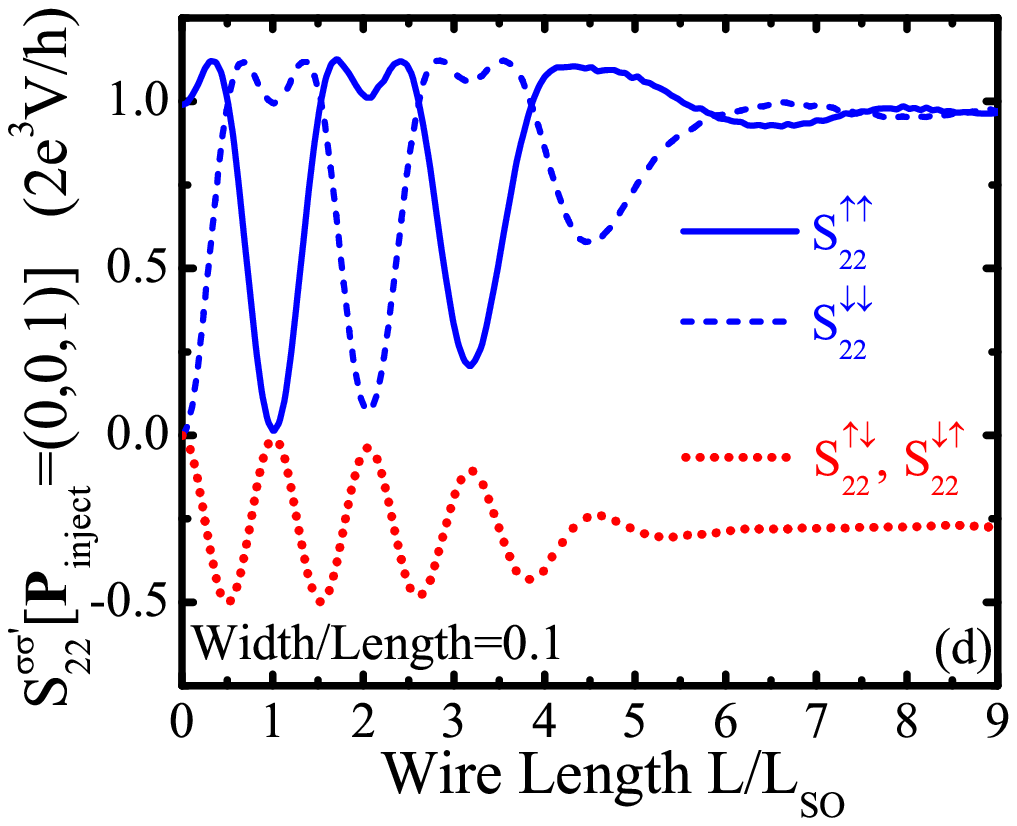,scale=0.6,angle=0}}
\centerline{\psfig{file=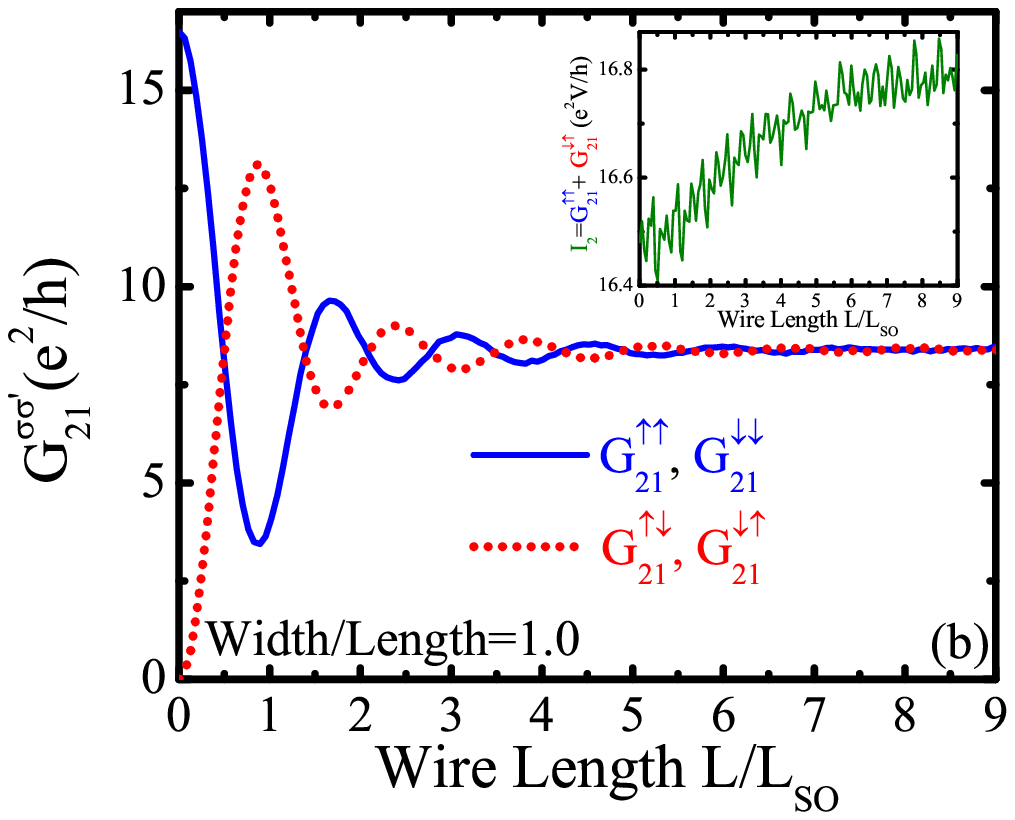,scale=0.6,angle=0} \hspace{0.5in} \psfig{file=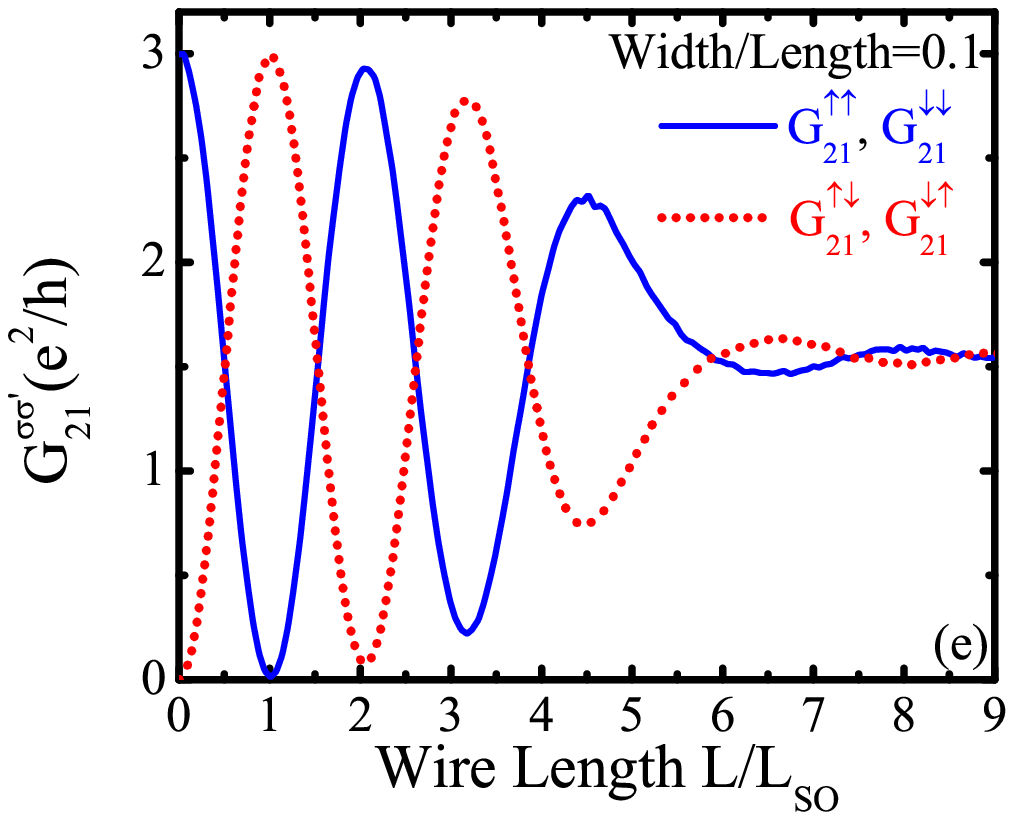,scale=0.6,angle=0}}
\centerline{\psfig{file=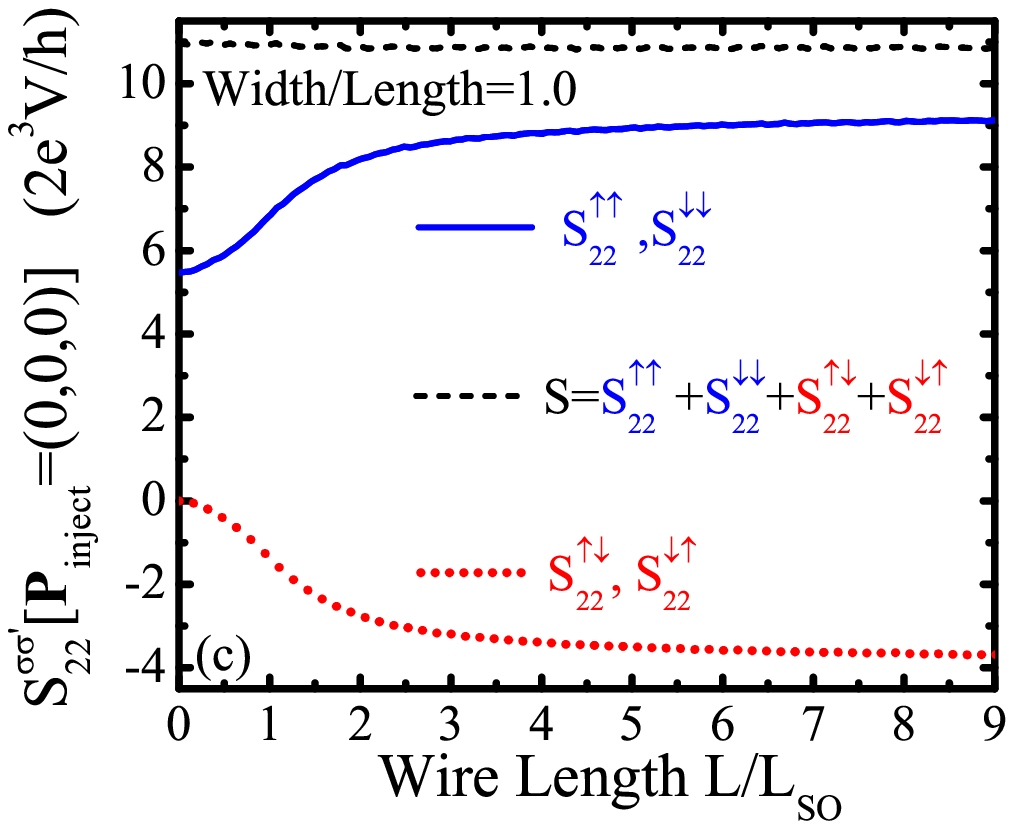,scale=0.6,angle=0} \hspace{0.5in} \psfig{file=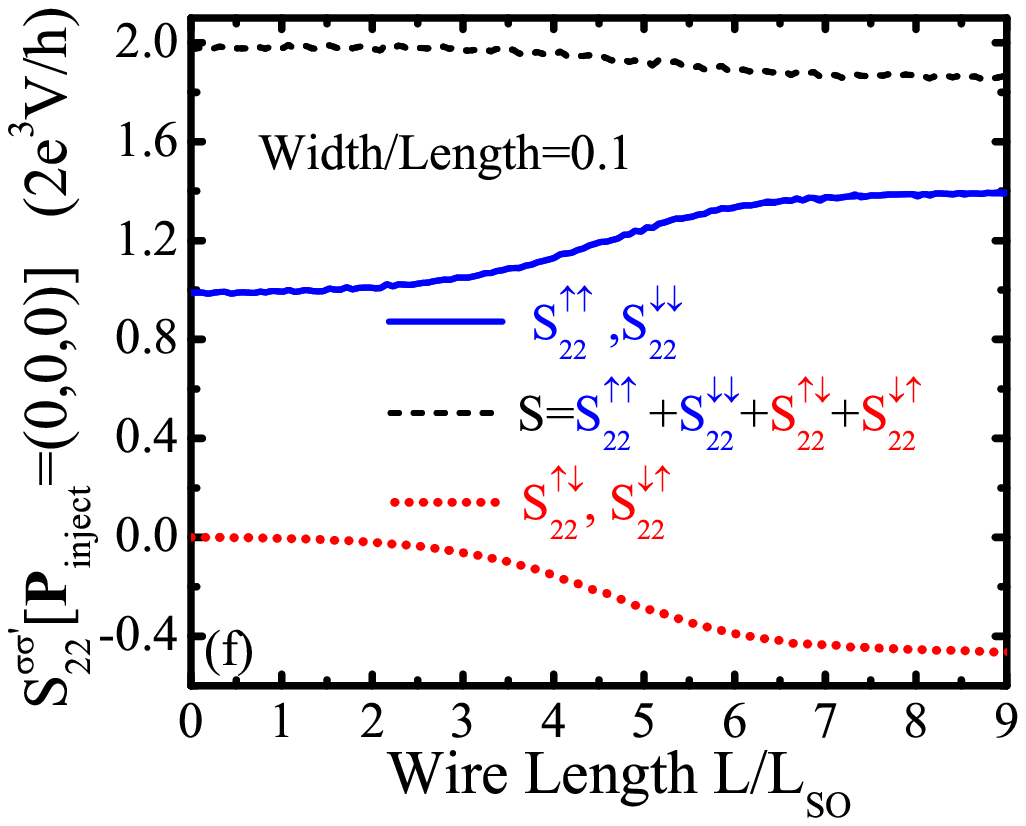,scale=0.6,angle=0}}
\caption{(Color online) Zero-frequency spin-resolved shot noise power $S_{22}^{\sigma \sigma^\prime}$ [panels (a), (d)] and spin-resolved conductances $G_{21}^{\sigma \sigma^\prime}$ [panels (b), (e)], which determined different Fano factors in Fig.~\ref{fig:noise}, for current detected in the right lead after the injection of spin-polarized (along the $z$-axis) charge current from the left lead into the diffusive wire with the Rashba SO coupling of strength $L/L_{\rm SO}$. The quantum wire is wide in panels (a), (b) and narrow in panels (d), (e). The inset in panel (b) shows weak antilocalization enhanced detected current in the right lead $I_2=I^\uparrow + I^\downarrow$ of ferromagnet/SO-coupled-wire/paramagnet setup. The spin-resolved noise for unpolarized current injection is shown 
in panels (e) and (f), whose sums give limiting curves (for $|{\bf P}_{\rm inject}|=0$) in Fig.~\ref{fig:3d}.}\label{fig:spin_resolved}
\end{figure*}

\section{Fano factor as quantifier of transported spin coherence} \label{sec:fano}

To understand the evolution of quantum coherence of transported spin, we use fully quantum transport formalism~\cite{nikolic_purity} which treats both the spin and  orbital dynamics phase coherently. 
This allows us to obtain the spin density matrix of charge current in the right lead in terms of the same spin-resolved transmission matrix ${\bf t}_{21}^{\sigma \sigma'}$ used to obtain the spin-resolved shot noise power $S_{22}^{\sigma \sigma^\prime}$. Note that traditional description of DP spin dephasing treats charge propagation semiclassically while the dynamics of its spin is described via quantum evolution of the 
spin density matrix.~\cite{wl_spin1}

For example, if a spin-$\uparrow$ polarized electron is injected from  the left lead through a conducting channel $|{\rm in}\rangle \equiv | m \rangle \otimes |\uparrow \rangle$, a pure state emerging in the right lead is described by the linear combination of the outgoing channels, $|{\rm out} \rangle = \sum_{n \sigma} 
[{\bf t}_{2 1}^{\sigma \uparrow}]_{nm} |n \rangle \otimes |\sigma \rangle$. Such {\em non-separable} state encodes entanglement of spin to orbital conducting channels, which is the source of 
{\em spin decoherence}~\cite{zeh_book} since the spin density matrix obtained by tracing over the orbital transverse propagating states $|n \rangle$ in the right lead will have the Bloch vector $|{\bf P}_{\rm detect}|<1$. Further decrease in the value of $|{\bf P}_{\rm detect}|$ is generated by {\em spin dephasing}~\cite{zeh_book} due to averaging over all orbital incoming channels $|m\rangle$ to produce the final spin density matrix of the detected charge current in the right lead~\cite{nikolic_purity}
\begin{widetext}
\begin{equation} \label{eq:rho_c}
\hat{\rho}^\uparrow_{\rm detect} =  \frac{e^2/h}{G^{\uparrow \uparrow}_{21} + G^{\downarrow \uparrow}_{21}} \! \sum_{n,m=1}^M \!\!\!
\left( \begin{array}{cc}
     |[{\bf t}_{2 1}^{\uparrow \uparrow}]_{nm}|^2 &  [{\bf t}_{2 1}^{\uparrow \uparrow}]_{nm}
       [{\bf t}_{21}^{\downarrow \uparrow}]_{nm}^*  \\ 
       
      [{\bf t}^{\uparrow \uparrow}_{21}]_{nm}^* [{\bf t}^{\downarrow \uparrow}_{21}]_{nm} &  
      |[{\bf t}_{21}^{\downarrow\uparrow}]_{nm}|^2
  \end{array} \right) = \frac{1}{2} \left( 1+ {\bf P}_{\rm detect} \cdot \hat{\bm \sigma} \right),
\end{equation}
\end{widetext}
and experimentally measurable Bloch vector ${\bf P}_{\rm detect}$ associated with it.
As demonstrated in Fig.~\ref{fig:noise}(c), in narrow wires quantum coherence of transported spin measured by the Bloch vector $|{\bf P}_{\rm detect}|$ extracted from $\hat{\rho}^\uparrow_{\rm detect}$ in Eq.~(\ref{eq:rho_c}) remains close to one for $L \lesssim L_{\rm SO}$. The preservation of quantum coherence also allows for spin-interference signatures to become visible in the shot noise of Fig.~\ref{fig:noise}(a) as ``Rabi oscillations'' of the Fano factor between $F_{\sigma \rightarrow \sigma'}=1/3$ and $F_{\sigma \rightarrow \sigma'}=1$ on the $L_{\rm SO}$-scale. 

\section{Discussion} \label{sec:discussion}

The phenomenological model of Ref.~\onlinecite{lamacraft2004}, characterized by the spin-relaxation length 
$L_s$ (which in the bulk SO coupled systems with weak disorder is identical~\cite{zutic2004,wl_spin1} to 
the spin precession length $L_{\rm SO}$),  finds $F_{\uparrow \rightarrow \uparrow\downarrow}(L \gg L_s)=2/3$. This in contrast to our $F_{\uparrow \rightarrow \uparrow\downarrow}(L \gg L_{\rm SO})=0.55$ governed by the parameters of microscopic Rashba Hamiltonian where further reduction of $F_{\uparrow \rightarrow \uparrow\downarrow}(L \gg L_s) 
< 0.55$ is induced by the geometrical confinement effects increasing spin coherence. 

As regards the spin-valve setups, the semislassical (Boltzmann-Langevine) approach to spin-dependent shot noise employed in Ref.~\onlinecite{mishchenko2003} predicts Fano factors $F_{\uparrow \rightarrow \uparrow}(L \gg L_s) = F_{\uparrow \rightarrow \downarrow}(L \gg L_s) = 1/3$ for arbitrary microscopic spin relaxation processes within the normal region, while we find $F_{\uparrow \rightarrow \uparrow}(L \gg L_{\rm SO}) = F_{\uparrow \rightarrow \downarrow}(L \gg L_{\rm SO}) \simeq 0.7$ for (wide) Rashba SO coupled wires. Furthermore, oscillatory behavior 
of the Fano factor versus $L/L_{\rm SO}$ exhibited in our Fig.~\ref{fig:noise}, especially conspicuous when 
quantum coherence of (partially coherent $0<|{\bf P}| < 1$) spin is increased in narrow wires, can emerge 
from approach of Ref.~\onlinecite{mishchenko2003} where spin dynamics is characterized only by $L_s$ (being 
much larger than mean free path with no restrictions imposed on its relation to the system size) rather than 
by the full spin density matrix.

To elucidate the source of these apparent discrepancies, we provide in Fig.~\ref{fig:spin_resolved} 
detailed microscopic picture of auto- and cross-correlations between spin resolved currents, as well 
as of spin-resolved conductances, which define our Fano factors at different SO coupling strengths 
$L/L_{\rm SO}$. It is obvious oscillations of both $S_{22}^{\sigma \sigma^\prime}$ and  $G_{21}^{\sigma \sigma^\prime}$ due to partially coherent spin precession, visible as long as $|{\bf P}| > 0$, can be captured 
only through fully quantum treatment of both spin dynamics and charge propagation (where spin memory between 
successive scattering events is taken account~\cite{pareek2002}). As regards the asymptotic values $F_{\uparrow \rightarrow \uparrow}(L \gg L_{\rm SO})$, this is determined by the shot noise $S_{22}^{\uparrow \uparrow}$ that 
is similar in both $L \ll L_{\rm SO}$ and $L \gg L_{\rm SO}$ limits, as well as by the current $I_2^\uparrow = G_{21}^{\uparrow\uparrow}V(L \gg L_{\rm SO})$ being half of its value at vanishing SO coupling $L/L_{\rm SO} \rightarrow 0$ [Fig.~\ref{fig:spin_resolved}(b),(e)]. This is due to the fact that at the exit of normal region 
with $L/L_{\rm SO} \gg 1$ charge current is unpolarized so that one of its spin subsystems if completely 
reflected from the detecting spin-selective (``analyzer'') electrode. 

Figure~\ref{fig:spin_resolved} also reveals that unpolarized charge current flowing out of the Rashba SO coupled 
region, after injected fully spin-polarized current was completely dephased $|{\bf P}|_{\rm detect}=0$ along 
the Rashba wire, still contains non-trivial cross-correlations between spin-resolved currents, as encoded in $S_{22}^{\uparrow \downarrow} = S_{22}^{\downarrow \uparrow} \neq 0$. They reduce $S_{22}^{\uparrow \downarrow} = S_{22}^{\downarrow \uparrow} < 0$ our Fano factor $F_{\uparrow \rightarrow \uparrow\downarrow}(L \gg L_s)= 0.55$  below $F_{\uparrow \rightarrow \uparrow\downarrow}(L \gg L_s)=2/3$ of Ref.~\onlinecite{lamacraft2004} (which we get approximately if we characterize the shot noise in the right lead only with  $S_{22}^{\uparrow \uparrow} + S_{22}^{\downarrow \downarrow}$).

\section{Concluding remarks} \label{sec:conclusion}

In conclusion, we have derived a scattering theory formula for the shot noise of 
charge and spin currents which takes as an input the degree of quantum coherence of 
injected spins $|{\bf P}_{\rm inject}|$. The application of this formalism to two-terminal 
multichannel diffusive quantum wires with the Rashba SO coupling unveils how spin decoherence 
and spin-dephasing processes  are essential  for the dramatic enhancement of the shot noise 
of charge current in spin-dependent  transport. That is, in narrow wires where the loss 
of spin coherence is suppressed and $|{\bf P}_{\rm detect}|$ decays much slower [Fig.~\ref{fig:noise}(c),(d)]  than in bulk systems, the enhancement of the Fano factor (above $F=1/3$ of spin-degenerate diffusive transport~\cite{beenakker1992}) in the strong SO coupling regime $L \gg L_{\rm SO}$ inducing fast spin 
dynamics within the wire is reduced. This occurs despite the fact that {\em partially coherent} 
$0<|{\bf P}_{\rm detect}|<1$ spin state continues to ``flip'', but through (partially coherent~\cite{nikolic_purity}) spin precession. To obtain the Fano factor of charge currents comprised of partially coherent spins requires 
to treat both charge propagation and spin dynamics quantum mechanically, as suggested by spin resolved shot noises and conductances in Fig.~\ref{fig:spin_resolved} (which cannot be reproduced by semiclassical approaches to spin-dependent shot noise where spin dynamics is captured only through generic spin-flip length~\cite{mishchenko2003}). Finally, a remarkable one-to-one correspondence 
between  the values of $F_{\uparrow \rightarrow \uparrow \downarrow}$ and the degree of quantum coherence 
$|{\bf P}_{\rm detect}|$ that we predict in Fig.~\ref{fig:noise}(e) offers exciting possibility to measure 
the coherence properties of transported spin in a {\em purely charge transport experiment} on an open SO-coupled structure, thereby offering an {\em all-electrical} alternative to currently employed optical tools to probe 
transport of spin coherence in semiconductors.~\cite{holleitner2006a}

\acknowledgments
This study was supported in part by the University of Delaware Research Foundation.



\begin{thebibliography}{1}

\bibitem{blanter2000} Ya.~M.~Blanter and M.~B\" uttiker, Phys. Rep. {\bf 336}, 1 (2000).

\bibitem{schomerus} H. Schomerus and P. Jacquod, J. Phys. A: Math. Gen. {\bf 38}, 10663 (2005).

\bibitem{steinbach1996} A.~H.~Steinbach, J.~M.~Martinis, and M.~H.~Devoret, Phys. Rev. Lett. {\bf 76}, 3806 (1996).

\bibitem{leo} E. Onac, F. Balestro, B. Trauzettel, C. F. J. Lodewijk, and L. P. Kouwenhoven, Phys. Rev. Lett. {\bf 96}, 026803 (2006).


\bibitem{beenakker1992} C. W. J. Beenakker and M. B\" uttiker, Phys. Rev. B {\bf 46}, 1889 (1992); K. E. Nagaev, Phys. Lett. A {\bf 169}, 103 (1992).

\bibitem{tserkovnyak2001} Y.~Tserkovnyak and A.~Brataas,  Phys. Rev. B {\bf 64}, 214402 (2001).

\bibitem{mishchenko2003} E.~G.~Mishchenko, Phys. Rev. B {\bf 68}, 100409(R) (2003).

\bibitem{lamacraft2004} A.~Lamacraft, Phys. Rev. B {\bf 69}, 081301(R) (2004).

\bibitem{nagaev2006} K.~E.~Nagaev and L.~I.~Glazman, Phys. Rev. B {\bf 73}, 054423 (2006).

\bibitem{belzig2004} W.~Belzig and M.~Zareyan, Phys. Rev. B {\bf 69}, 140407(R) (2004); M. Zareyan and 
W. Belzig, Europhys. Lett. {\bf 70}, 817 (2005).

\bibitem{cottet2004}  A. Cottet, W. Belzig, and C. Bruder, Phys. Rev. Lett. {\bf 92}, 206801 (2004).
 
\bibitem{egues2005} J.~C.~Egues, G.~Burkard, and D. Loss, Phys. Rev. Lett. {\bf 89}, 176401 (2002); 
J. C. Egues, G. Burkard, D. S. Saraga, J. Schliemann, and D. Loss, Phys. Rev. B {\bf 72}, 235326 (2005).

\bibitem{beenakker1998} C. W. J. Beenakker, Phys. Rev. Lett. {\bf 81}, 1829 (1998).

\bibitem{winkler_book} R. Winkler, {\em Spin-Orbit Coupling Effects in Two-Dimensional Electron and Hole 
Systems} (Springer, Berlin, 2003). 

\bibitem{zutic2004} I. \v Zuti\' c, J. Fabian, and S. Das Sarma, Rev. Mod. Phys. {\bf 76}, 323 (2004).

\bibitem{ossipov2006} A. Ossipov, J. H. Bardarson, C. W. J. Beenakker, J. Tworzyd\l o, and M. Titov, cond-mat/0603524.

\bibitem{nikolic_purity} B.~K.~Nikoli\' c and S.~Souma, Phys. Rev. B {\bf 71}, 195328  (2005). 

\bibitem{wl_spin1} A.~G.~Mal'shukov and K.~A.~Chao, Phys. Rev. B {\bf 61}, R2413 (2000); A.~A.~Kiselev and K.~W.~Kim, Phys. Rev. B {\bf 61}, 13115 (2000).

\bibitem{buttiker1992a} M.~B\" uttiker, Phys. Rev. B {\bf 46}, 12485 (1992).

\bibitem{holleitner2006a} A. W. Holleitner, V. Sih, R. C. Myers, A. C. Gossard, and D. D. Awschalom, Phys. Rev. Lett. {\bf 97}, 036805 (2006).

\bibitem{sauret2004} O.~Sauret and D.~Feinberg, Phys. Rev. Lett. {\bf 92}, 106601 (2004).

\bibitem{aleiner2001} I. L. Aleiner and V. I. Fal'ko, Phys. Rev. Lett. {\bf 87}, 256801 (2001).

\bibitem{pareek2002} T. P. Pareek and P. Bruno, Phys. Rev. B {\bf 65}, 241305(R) (2002).

\bibitem{zeh_book} E.~Joos, H.~D.~Zeh, C.~Kiefer, D.~Giulini, J.~Kupsch, and I.-O.~Stamatescu, {\em Decoherence and the appearance of a classical world in quantum theory} (2nd ed., Springer-Verlag, Heidelberg, 2003).


\end{thebibliography}
\end{document}